\documentclass[10pt,journal,compsoc]{IEEEtran}
\usepackage{cite}
\usepackage{amsmath,amssymb,amsfonts}
\usepackage{algorithmic}
\usepackage{graphicx}
\usepackage{textcomp}
\usepackage{caption}
\usepackage{subcaption}
\usepackage{balance}

\usepackage{array}
\usepackage{dblfloatfix}
\usepackage{float}
\usepackage{algorithm}
\usepackage[underline=true]{pgf-umlsd}
\usetikzlibrary{calc}
\usepackage{listings}
\usepackage{xcolor}
\def\BibTeX{{\rm B\kern-.05em{\sc i\kern-.025em b}\kern-.08em
    T\kern-.1667em\lower.7ex\hbox{E}\kern-.125emX}}
    
\usepackage{xcolor}
\definecolor{backcolour}{RGB}{238,238,238}
\lstdefinestyle{mystyle}{
    backgroundcolor=\color{backcolour}  
}
\begin{document}

    \title{Identity-Based Authentication for On-Demand Charging of Electric Vehicles\\
    {\footnotesize }
    }
    
    \author{Surudhi~Asokraj\textsuperscript{\textdagger},
        Tommaso~Bianchi\textsuperscript{\textdagger},
        Alessandro~Brighente,
        Mauro~Conti,
        Radha~Poovendran
    \IEEEcompsocitemizethanks{\IEEEcompsocthanksitem S. Asokraj and R. Poovendran are with the Network Security Lab at the Department of Electrical and Computer Engineering, University of Washington, Seattle, WA 98195-2500, USA. \protect\\
    E-mail: \IEEEauthorblockA{ \{surudh22, rp3\}@uw.edu}
    \IEEEcompsocthanksitem T. Bianchi, A. Brighente and M. Conti are with the Department of Mathematics, University of Padua, Via Trieste 63, Padua, 35121, Italy.\protect\\
    E-mail: \IEEEauthorblockA{tommaso.bianchi@studenti.unipd.it, \{alessandro.brighente, mauro.conti\}@unipd.it} \\
    
    \IEEEcompsocthanksitem \textdagger: These authors contributed equally}
    
     \thanks{This work has been submitted to the IEEE for possible publication. Copyright may be transferred without notice, after which this version may no longer be accessible.}
    }
    
    
     
    
    
    \IEEEtitleabstractindextext{
        \begin{abstract}
Dynamic wireless power transfer provides means for charging Electric Vehicles (EVs) while driving, avoiding stopping for charging and hence fostering their widespread adoption. Researchers devoted much effort over the last decade to provide a reliable infrastructure for potential users to improve comfort and time management. Due to the severe security and performance system requirements, the different scheme proposed in last years lack of a unified protocol involving the modern architecture model with merged authentication and billing processes. Furthermore, they require the continuous interaction of the trusted entity during the process, increasing the delay for the communication and reducing security due to the large number of message exchanges. In this paper, we propose a secure, computationally lightweight, unified protocol for fast authentication and billing that provides on-demand dynamic charging to comprehensively deal with all the computational and security constraints. The protocol employs an ID-based public encryption scheme to manage mutual authentication and pseudonyms to preserve the user's identity across multiple charging processes. Compared to state-of-the-art authentication protocols, our proposal overcomes the problem of overwhelming interactions and provides public scheme security against the use of simple operations in wide open communications without impacting on performance.
\end{abstract}


    }
    
    \maketitle

    \IEEEraisesectionheading{\section{Introduction}\label{sec:introduction}}

\IEEEPARstart{O}{ver} the last few years, we have seen a rapid evolution in Electric Vehicles (EVs) due to technological advancements and the environmental awareness of the population in terms of Carbon Dioxide ($CO_2$) emissions. This realization of the necessity of resource-efficient choices led to a transition to eco-friendly automobiles against the traditional fossil fuel engine vehicles. Various countries encourage the use of electric-powered technologies to reduce $CO_2$ emissions by 2050 \cite{energy}. The increase in sales and use of EVs need to be accompanied by the development of adequate infrastructure to match the surge in battery charging demand\cite{battery}. Initially, wired or wireless static charging systems were proposed, with charging stations installed along roads to allow drivers to stop if needed\cite{plugged-in}. The downside to this approach is that it requires the vehicle to halt for the charging, and the stop usually requires a considerable amount of time \cite{pareek}. Due to the aforementioned problem and the currently limited battery autonomy of an EV, a capillary installation of charging stations is necessary to give the possibility to charge the vehicle during long-distance trips without running out of charge. These issues might impact the customer's quality of service by requiring frequent stops and increasing the total time needed for a journey. \par 

Dynamic wireless power transfer systems are a promising technology that charges the battery using wireless induction while the vehicle is moving \cite{suh-kim}. This technology employs the new concept of On-line Electric Vehicle (OLEV) \cite{olev}, where the power-line infrastructure installed beneath the road is responsible for charging the EVs. The model entails installing charging pads under the road, which provide energy to the EV and aid in billing the right customer. In this scenario, the vehicles need to communicate with different infrastructure entities, requiring the deployment of a heterogeneous group of technologies and systems. \par

The variety of entities involved in the charging process and the vehicle's connection to the network make it necessary to protect the users from threats against security and privacy. The absence of such defenses could lead to the tracking of the customer activities or the possibility for a malevolent user to get free charging at the expense of another legitimate EV. Since the vehicle moves at high speeds, the system needs an appropriate authorization and mutual authentication scheme. In this way, the customer using the system needs to pay for the service without the provider knowing the real identity of the vehicle owner, thus dictating the necessity of a secure protocol addressing these issues. Moreover, the dynamic charging infrastructure imposes constraints in terms of computation and communication efficiency that impact the scheme design, which must be fast and lightweight. Furthermore, the billing must be fair and verifiable by both entities (customer and provider) to guarantee reliability. If there is a conflict between the two parties, there must be a way to verify the charging process and resolve the dispute. The tamper-proof On-Board Unit (OBU), mounted on the EV, usually provides this function. \par

The high number of different requirements needed for this system can deter the implementation of the infrastructure \cite{machura-critical}. In fact, without a carefully designed system, it is possible to incur weighting the scheme and introduce security vulnerabilities. However, researchers and federal transportation agencies have done much work to build a secure and safe environment for dynamic charging \cite{hutchinson}. Many studies are trying to implement and test lanes for dynamic wireless charging in different locations with promising results \cite{road-example}. The research for a secure mechanism in the authentication and billing of the OLEV has different implementations and conclusions but with other goals, as we describe in the next section. Our approach wants to address all the issues that the dynamic charging system presents, using the most recent model for the network and implementing a public-key cryptosystem that respects the limits imposed by the infrastructure. In this concern, we propose a scheme to overcome the in-advance energy purchasing by addressing the authentication and billing phases without losing performance respect to the SOTA protocol. In our system, the privacy of the user is preserved without sending any critical information in clear, taking advantage of the identity-based public key infrastructure.
Following are the main contribution of our work:

\begin{itemize}
    \item By generating pseudonyms based on ID-based public key infrastructure and the Boneh-Franklin system \cite{boneh-franklin}, we provide the security of public-key encryption and key derivation for the session. Resulting computational cost is comparable with protocols that involve a similar infrastructure model and entities\cite{babu}.
    \item We implement On-Demand charging request to provide the possibility to stop the charging process on user decision, without the need to specify the amount of power required at the beginning of the process.
    \item Our approach includes a secure billing scheme without weighing down on the protocol, maintaining performances comparable with the SOTA solutions.
    \item We show that the proposed protocol is robust and secure against a variety of attacks, particularly against \textit{free-riding} and  \textit{double-spending} (\cite{zhao, rabieh}).
\end{itemize} \par

The rest of the paper is structured as follows: in Section II, we discuss the work that has been done in the dynamic wireless charging area.In Section III, we describe the network and the adversary models, while Section IV explains the mathematical tools used. Sections V and VI are dedicated to the description of the protocol. Subsequent sections VII and VIII contain the security and performance analysis, respectively. We provide a brief discussion in Section X and draw our conclusions in Section XI.

    
    \section{Related Work}

With the advent of dynamic wireless charging in the Vehicular Ad hoc NETwork (VANET) scenario, many different studies have proposed secure and reliable schemes for the users of a dynamic wireless charging service\cite{hussain2015, li, hussain2017, zhao, rabieh, roman, hamouid, elghanam, babu}. The effort is focused on finding a lightweight, preferably mutual, authentication and billing scheme. Over time, the system model has evolved from a simple central organization directly managing the Charging Pads (CP) to a more complex cloud environment. The most significant contributions in the last years includes the scheme proposed by Hussain \textit{et al.} \cite{hussain2015}, presenting a privacy-aware and bidirectional mechanism for Dynamic Wireless Power Transfer (DWPT). The model is simple, with a central authority directly interacting with the EVs and the CPs. To preserve user privacy during EV authentication, they use pseudonyms for the vehicles and a hash chain-based mechanism. This simple model approach using hashes does not fit the needs of a cloud-based environment, so other models are explored. The later models for the service provider infrastructure comprises of the Charging Service Provider Authority (CSPA) with multiple Road-Side Units (RSU) that provide control and computational power at the network edge. These units are controlled by a Fog Server (FS), and usually involves other entities like banks or trusted authorities. \par

In 2017, Li \textit{et al.} \cite{li} developed a DWPT system addressing authentication and privacy protection. The author recognizes that the scheme is not optimized for the charging process, but it is simple, robust, privacy-preserving, and scalable. The scheme relies on symmetric keys based on pre-distribution and the Spatio-temporal location of the EV. The EV and the CPs exchange messages encrypted with the symmetric keys, one for each charging pad. Different keys result in high overhead, devoting a significant part of the communication to the authentication process. In 2017, Hussain \textit{et al.} \cite{hussain2017} proposed a new version of their scheme, addressing the billing part but using the same authentication and privacy-preserving mechanisms. The same model comprises a charging service authority and the charging pads. Hussain \textit{et al.} used the hash-chain for the authentication between the EV and CPs.\par

Zhao \textit{et al.} \cite{zhao} presented a scheme based on a Registration Authority (RA) in charge of generating all the parameters for public-key encryption and signing. Other entities, such as banks, provide authentication tokens for the users. The main objective of the protocol is to be secure against free-riding \cite{zhao}, an attack in the VANET scenario where an EV tries to get an electric charge without paying. They address this issue by periodically checking the battery level of a vehicle and verifying that it is not increasing during a non-authenticated period. In \cite{rabieh}, Rabieh \textit{et al.} developed a scheme based only on blind signatures, hash chains, and XOR functions where the EV needs to buy tickets for the charging process in advance and offline. The protocol provides authentication, user privacy and anonymity. Additionally, it addresses the problem of double-spending, another specific threat in the VANET scenario where the EV tries to use an old transaction to get charged. An authentication protocol for a cloud environment is proposed in \cite{roman}, with a hierarchical architecture comprising a fog server. The scheme is based on short signatures, blind signatures and bilinear pairing. The EV needs to be registered with a charging company and purchase tickets before charging. After the purchase, the EV can proceed with the charging request. \par

Hamouid \textit{et al.} \cite{hamouid} proposed a protocol that uses verifiable encryption, authenticated pairwise keys and coin hash chains, addressing privacy through anonymous authentication with pseudonyms. After registering and generating a token and coin chain for validating the charging process, the scheme does not involve the charging service provider during the authentication between EV and CP. ElGhanam \textit{et al.} \cite{elghanam} presented a scheme accounting for authentication and billing processes. Here, the Charging Service Company (CSC) is considered the trusted party, while different Pad Owners (PO) own the charging segments composed of different pads. The EV registers and authenticates with the CSC. After the authorization, it can authenticate with the PO and start the charging process using a hash chain. The protocol relies on symmetric and asymmetric encryption schemes and elliptic curves for signing the messages. The scheme is robust against a variety of attacks and it is considered lightweight even if many messages are exchanged in real-time during the charging process. The authors in \cite{babu} proposed a protocol involving only the authentication part. The scheme works with elliptic curve cryptography for signing, and after a system initialization phase, the fog servers and the vehicles are registered to the Company Charger Server. After that, when EV needs to charge, it can initiate mutual authentication. In this phase, EV exchanges encrypted messages with FS and RSU, using a hash chain to provide authenticity with the CPs. The protocol's security is verified with the help of the Scyther tool and a MIRACL-based testbed experiment, proving the lightweight computational cost of the proposed scheme. \par

In previous works, there is usually a lack of mutual authentication. If it is accounted for, the communication overhead is high due to the number of messages exchanged with multiple entities. Another problem is that the service provider is assumed to be a trusted authority in most cases, which is not always the case. Moreover, the proposed protocols do not consider the possibility to start and stop the charging process on-demand, according to the user's needs. The schemes mentioned above need the EV to decide in advance the amount of money or charge it wants for the subsequent iteration of the protocol. In this context, we provide an authentication protocol scheme addressing mutual authentication and on-demand charging for the EV, keeping the robustness against various types of attack and preserving the user's privacy using ID-based public key cryptography and hash chains through pseudonyms.
    
    \begin{figure*}[htb]
    \centering
    \includegraphics[width=15cm]{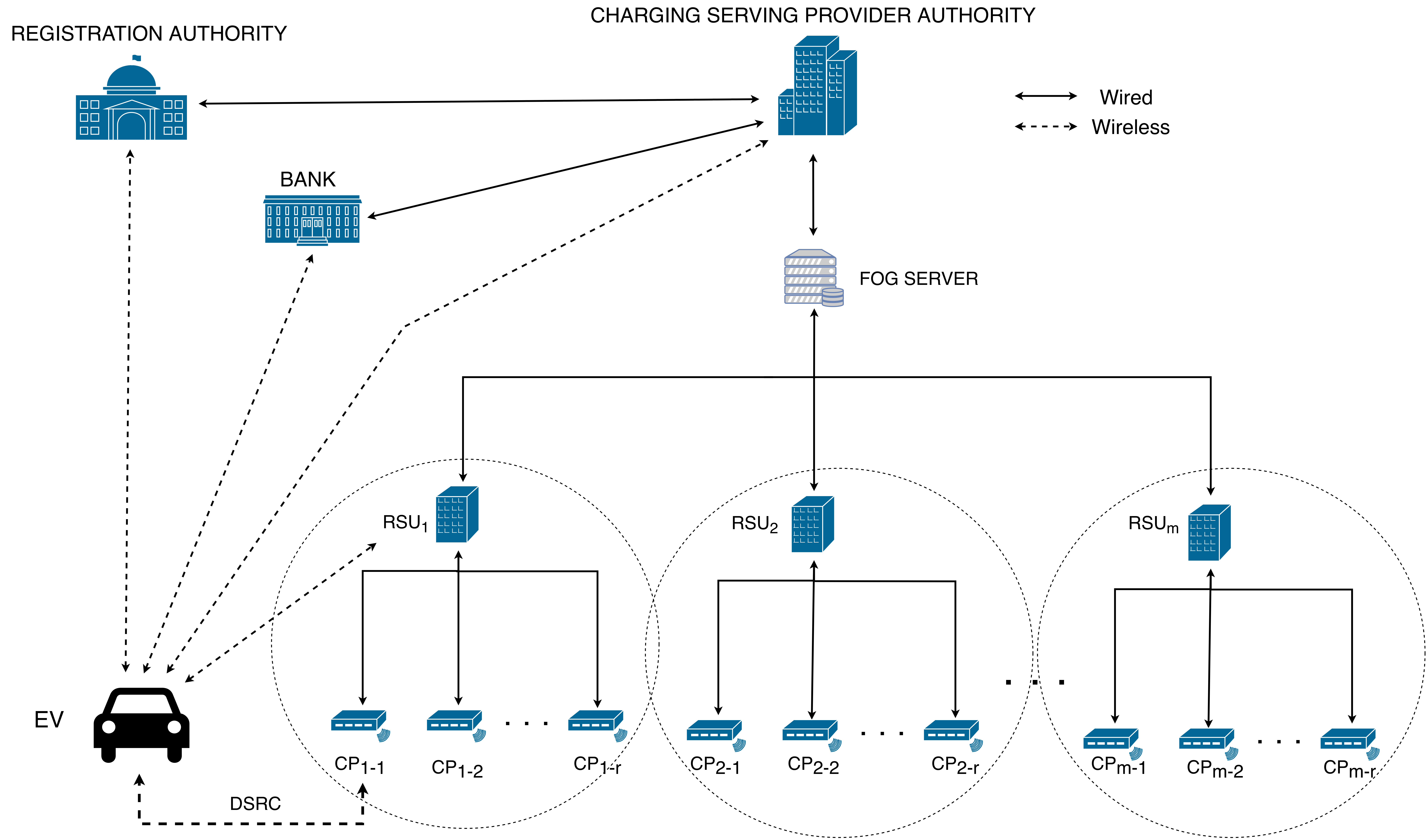}
    \caption{Network Scheme of the proposed model, illustrating connections between the various components of the service provider, the electric vehicle and the external entities.}
    \label{fig:model_scheme}
\end{figure*}
    
    \section{Network and Adversary Model}
We present the network model in Section III.A and the adversary model in Section III.B.

    \subsection{Network model} 
    The network is structured as depicted in Fig. \ref{fig:model_scheme}. A trusted certificate deploying authority, referenced as Registration Authority (\textit{RA}) in Fig. \ref{fig:model_scheme}, is in charge of publishing the ID-Based Public Key cryptography parameters. The Charging Service Provider Authority (\textit{CSPA}) is the organization providing the infrastructure for the charging system, such as the Road-Side Units (\textit{RSU}) and Charging Pads (\textit{CP}). All of the components in this charge service provider network are linked via a wired and reliable connections, as shown in Fig. \ref{fig:model_scheme}. The CSPA is structured as a cloud environment, using intermediary devices to reach the edge of the network. In the middle, there are Fog Servers (\textit{FS}), which in our scenario, are responsible for forwarding requests from the EV to CSPA and from the CSPA to RSUs during the authentication process. Each FS is in charge of \textit{m} RSUs. The RSUs are responsible for authentication between the EVs and the CPs and provide the necessary cryptographic elements for the charging process. The CPs are the last and more constrained part of the network, with the hardware necessary for the wireless charging of the EVs. They are disposed in line under the road, with little space between each CP. Each RSU is in charge only of a subset of the CPs, spreading the work across multiple pad segments (composed by up to \textit{r} CPs). The EV can communicate between RA, CSPA and RSU through a wireless channel, while it communicates with the CPs using a Dedicated Short Range Communication (\textit{DSRC}) channel.
    
    \subsection{Adversary Model}
    Our system considers the Dolev-Yao\cite{dolev-yao} model where an adversary can compose, replay, intercept and forge messages but without the correct cryptographic keys, he/she cannot decipher the messages. The adversary, in particular, desires to obtain a free charge, in which case they are referred to as \textit{free-riders}, or to pose as another authenticated user (\textit{double-spending}) by reusing an old pseudonym or token.  In our proposed scheme, we assume that only the RA is trustworthy and knows the real identity of the EVs, which should never be revealed during the communication. 
    \par In our model, we consider a cryptographic one-way hash function to provide strong collision resistance and pre-image resistance. The wired communication network is considered secure due to the use of a symmetric key mechanism, but an adversary can connect to the network and sniff the traffic. An adversary may be able to infer useful information that could be utilized for further attacks, such as \textit{free-riding} or replay attacks if the protocol is poorly constructed. As we describe in later sections, our scheme is secure against the most common attacks, and access to the wireless network provides no advantage to an adversary. Group keys, generated by the CSPA and known only to trusted parties, are used to encrypt traffic between different entities within the network.
    



\section{Preliminaries}
In this section we briefly describe the mathematical tools used in our protocol.
\subsection{Bilinear pairing} 
A bilinear pairing is defined on a bilinear group possessing a bilinear mapping with two multiplicative cyclic groups $\mathbb{G}_1, \mathbb{G}_2$ of finite order \textit{n}, a generator P of $\mathbb{G}_1$ and \textit{e} is a bilinear map $e: \mathbb{G}_1 \times \mathbb{G}_1 \rightarrow \mathbb{G}_2$.

$\mathbb{G}_1$ is said to be a bilinear group if group action in $\mathbb{G}_1$ can be computed efficiently and there exists both a group $\mathbb{G}_2$ and an efficiently computable bilinear map \textit{e}. It satisfies the bilinear property, where $e(aP,bQ) = e(P,Q)^{ab}$, for all $P,Q \in \mathbb{G}_1$ and all $a,b \in \mathbb{Z}$. Given two groups $\mathbb{G}_1$ and $\mathbb{G}_2$ of order \textit{q}, where \textit{q} is a large prime number, the bilinear map \textit{e} is:
\begin{itemize}
    \item non-degenerate and does not send all pairs in $\mathbb{G}_1 \times \mathbb{G}_1$ to the identity in $\mathbb{G}_2$. $\mathbb{G}_1, \mathbb{G}_2$ are groups of prime order and this implies that if \textit{P} is a generator of $\mathbb{G}_1$ then $e(P,P)$ is a generator of $\mathbb{G}_2$;
    \item computable such that there exists an efficient algorithm to compute $e(P,Q)$ for any $P,Q \in \mathbb{G}_1$.
\end{itemize} 

Decisional Diffie-Hellman in $\mathbb{G}_1$ is easier to solve and hence, not preferred to build the cryptosystem. We use a variant of the computational Diffie-Hellman assumption, Weil Diffie-Hellman (WDH) to guarantee the security of the ID-based encryption scheme \cite{boneh-franklin}.
According to WDH, it is hard, given $<P,aP,bP,cP>$ for some $a,b,c \in \mathbb{Z}^*_p$, to compute $e(P,P)^{abc} \in \mathbb{G}_2$.

\subsection{Hash functions}
In our protocol, we propose the use of four different hash functions that map the input in different output spaces.
\begin{enumerate}
    \item H1: maps $\{0,1\}^* \rightarrow \mathbb{G}_1$;
    \item H2: maps $\{0,1\}^* \rightarrow \mathbb{Z}^{*}_q$;
    \item H3: maps $\mathbb{G}_2 \rightarrow \{0,1\}^m$;
    \item \textit{h}: linear \textit{e-one-way} hash function $\mathbb{G}_1 \times \mathbb{Z}^*_q \rightarrow \mathbb{G}_1$.
\end{enumerate}
The last hash function is essential because it allows the computation of the private keys during the protocol between two parties that need to share a secret without transmitting it effectively. Especially, the \textit{e-one-way} hash function \textit{h} has two properties:
\begin{itemize}
    \item \textit{Commutative e-one-way hash property} declares that for all $P \in \mathbb{G}_1, a,x \in \mathbb{Z}^*_q, h(aP,x) = ah(P,x)$;
    \item The problem can not be solved by a probabilistic polynomial algorithm.
\end{itemize}
The first property is essential for the protocol to work during the computation of the pair of keys of the two involved parties in each session.

\subsection{ID-based public key encryption}
The cryptographic system used in our work is ID-based public-key cryptography. Our utilized system is built up upon pairing groups and elliptic curves \cite{boneh-franklin} on finite fields. A trusted party is responsible for publishing the parameters for the scheme, as the \textit{master public key}. This trusted agent generates a secret, called \textit{master secret key} that is used for private key derivation of the users who want to register to the service. 

\begin{figure}[b!]
    \includegraphics[width=8.5cm]{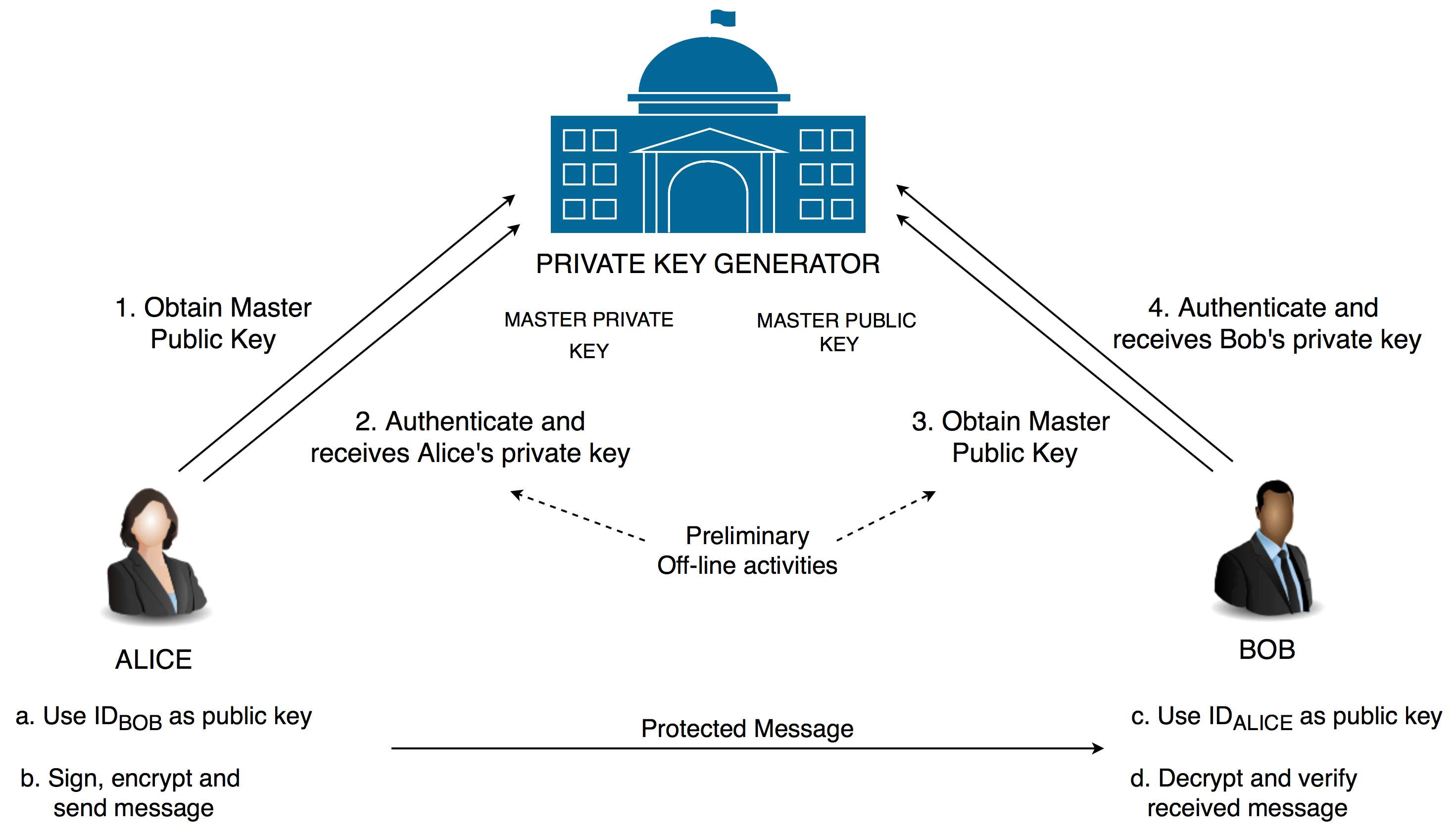}
    \caption{General scheme of ID-base encryption}
    \label{fig:id_enc}
\end{figure}

\par The distinction in our system is that the public key used is derived from a string or a parameter representing the user's identity, e.g., the email address, using the master public key. Instead, the private key is created by the trusted party using its secret and the entity's public key. The protocol uses bilinear pairing  to encrypt the message with the public key, and multiplication in the same group is sufficient for decryption. Fig. \ref{fig:id_enc} depicts a pictorial representation of what we discussed about ID-based encryption.

\subsection{Correctness}
In the proposed scheme, we use a set of hash functions and a bilinear map that, following some properties, allow two parties to compute the same result in the map using different parameters. The bilinear property of the bilinear map $e$ and the commutative property of the e-one-way hash function are crucial to understanding why the protocol works. \par
Given two entities, namely $E_1$ and $E_2$, we want to establish a session key, named differently for the two parties: $k_{E_1}$ and $k_{E_2}$. These keys lead to the same mathematical results thanks to the just cited properties. In general, with entities  $E_1$ and $E_2$, pseudonym $PID_{E_1}$ for $E_1$, RA as authority that publishes the public parameters, and \textit{ID} referring to the identity of the entity, we have:
\[ k_{E_1} \hspace{-2mm} = e(H_1(ID_{E_2}), h(sH_1(ID_{RA}), H_2(ID_{RA} \parallel PID^i_{E_1}))).\]

The first component belongs to $\mathbb{G}_1$ and it is multiplied by 1. The second component also belongs to $\mathbb{G}_1$ but the value inside the hash function contains the term \textit{s} belonging to $\mathbb{Z}^{*}_q$, multiplied by a value in $\mathbb{G}_1$, and the result of $H_2$ that belongs to $\mathbb{G}_1$. Therefore, we can apply the e-one-way function property: 
\[ k_{E_1} \hspace{-2mm} = e(H_1(ID_{E_2}), s h(H_1(ID_{RA}), H_2(ID_{RA} \parallel PID^i_{E_1}))).\]
 
Thus, having a $a=1$ and $b=s$, for the bilinear map property: 
\[ k_{E_1} \hspace{-2mm} = e(H_1(ID_{E_2}), h(H_1(ID_{RA}), H_2(ID_{RA} \parallel PID^i_{E_1})))^\textbf{s}.\]

The same applies for $k_{E_2}$ in a more direct way, having \textit{s} in the first component of the map. This indeed gives the same result, leading to $k_{E_1} = k_{E_2}$.
    
    \section{Proposed Authentication Protocol}
\begin{figure*}[t]
    \centering
    \includegraphics[width=18cm, height = 9cm]{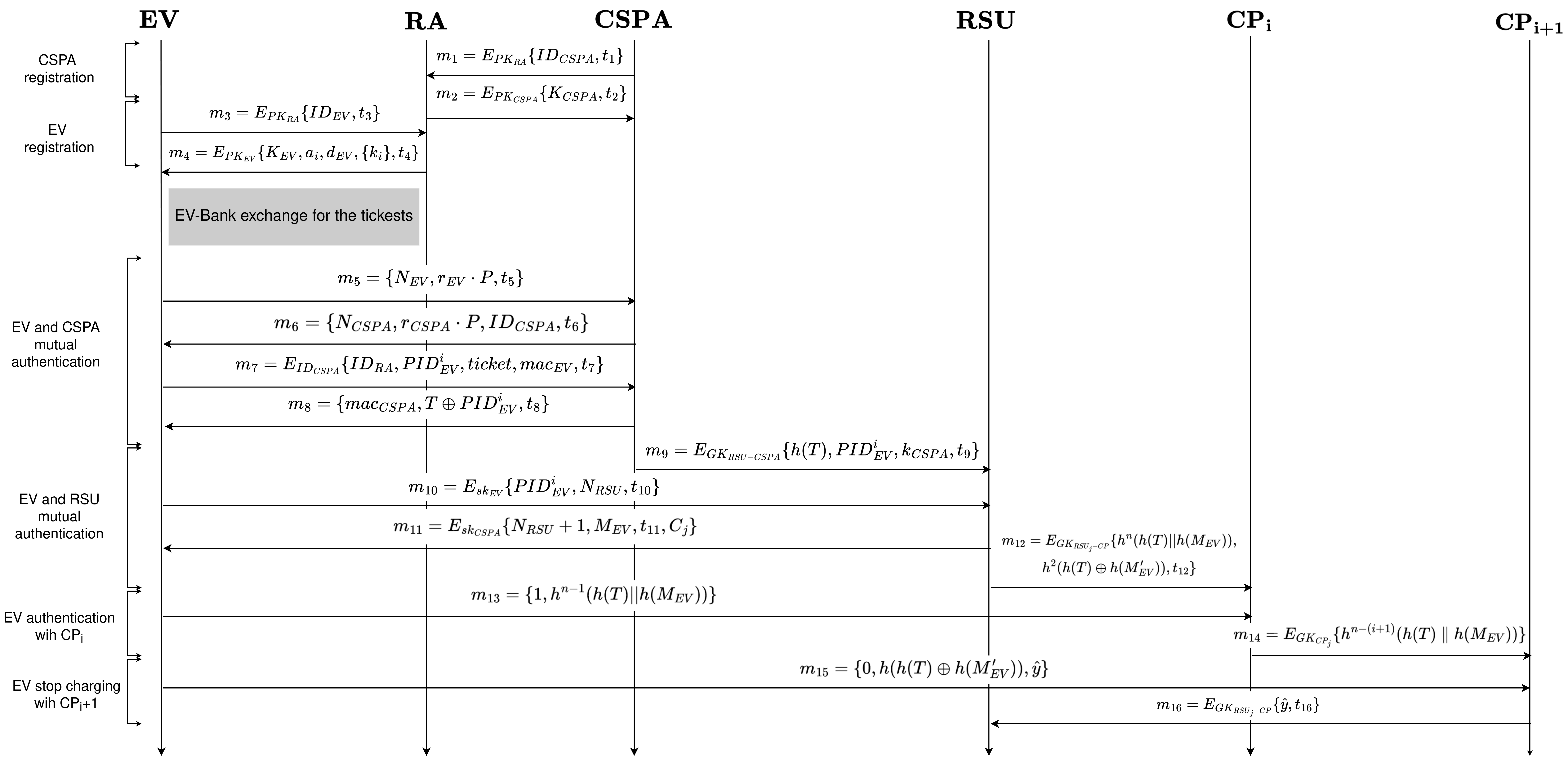}
    \caption{Sequence diagram of the authentication protocol}
    \label{fig:seq_diagram}
\end{figure*}

In Table \ref{table:symbols}, we provide the details of the notations used in the protocol. The protocol is based on the Boneh-Franklin identity-based public encryption scheme \cite{boneh-franklin}, adapting an authentication scheme meant for the mobile roaming environment by Wan \textit{et al.} \cite{wan}. The message exchange sequence during authentication between the different entities in the system is depicted in Fig. \ref{fig:seq_diagram}.

\subsection{System Initialization Phase}
    \begin{enumerate}
        \item The trusted Registration Authority (RA) generates and publishes the parameters for the ID-based Public Key Cryptography mechanism: the two multiplicative cyclic groups of finite order \textit{n}, the bilinear map \textit{e} with the generator \textit{P}, the public parameter \textit{sP}, its own ID and the hash functions $H_1$, $H_2$, $H_3$.
        \[<\mathbb{G}_1, \mathbb{G}_2, e, P, sP, ID_{RA}, H_1, H_2, H_3>\]
        In addition, RA chooses master secret $s \in \mathbb{Z}^*_q$.
        \item CSPA chooses group key for communication between CSPA and a RSU $\{GK_{RSU-CSPA}\}$ and sends it to the RSUs.
        \item CSPA also chooses group keys for communication between RSU and the CPs associated with it $\{GK_{RSU_j-CP}\}$ and delivers the respective one to the RSUs and CPs.
    \end{enumerate}
    
\begin{table}[t!]
\begin{center}
    \begin{tabular}{| p{2.5cm} p{5cm} |}
        \hline
        \textbf{SYMBOL} & \textbf{DESCRIPTION} \\
        \hline
        $EV$ & Electric Vehicle \\
        \hline
        $CSPA$ & Charge Serving Provider Authority \\
        \hline
        $RA$ & Registration Authority \\
        \hline
        $CP$ & Charging Pads \\
        \hline
        $RSU$ & Road Side Units \\
        \hline
        $ID_{EV}$, $ID_{RA}$ & Identities of EV and RA \\
        \hline
        $ID_{CSPA}$ & Identity of CSPA \\
        \hline
        $PK_{RA}, PK_{CSPA}$ & Public key of RA and CSPA \\
        \hline
        $GK_{RSU-CSPA}$ & Group key between RSU and CSPA \\
        \hline
        $GK_{RSU_j-CP}$ & Group key between RSU\textsubscript{j} and CPs \\
        \hline
        $GK_{CP_j}$ & Group key among CPs that belong to RSU\textsubscript{j} \\
        \hline
        $SK_{EV}, PK_{EV}$ & Private and Public key of EV \\
        \hline
        $k_{EV}, k_{CSPA}$ & ID-based secret key of EV and CSPA \\
        \hline
        $sk_{EV}, sk_{CSPA}$ & ID-based session key between EV and CSPA \\
        \hline
        $PID^{i}_{EV}$ & Pseudonym of EV \\
        \hline
        $N_{EV}, r_{EV}$ & Random nonces of EV \\
        \hline
        $N_{CSPA}, r_{CSPA}$ & Random nonces of CSPA\\
        \hline
        $E_{PK_{X}}(\cdot)$ & Asymmetric key encryption   \\
        \hline
        $E_{GK_{X-X}}(\cdot)$ & Symmetric encryption using group key \\
        \hline
        $h(\cdot), H(\cdot)$ & Cryptographic hash functions \\
        \hline
        $mac_{EV}, mac_{CSPA}$ & Message authentication code\\
        \hline
        $t$ & Timestamp \\
        \hline
        $T$ & Token \\
        \hline
    \end{tabular}
    \caption{Table of symbols.}
    \label{table:symbols}
\end{center}
\end{table}

\subsection{CSPA Registration Phase}
        \begin{enumerate}
            \item CSPA transmits its $ID_{CSPA}$ to RA along with timestamp $t_1$:
            \[CSPA \rightarrow RA: m_1 = E_{PK_{RA}}\{ID_{CSPA}, t_1\}.\]
            \item RA generates ID-based private key with the master secret and sends:
            \[RA \rightarrow CSPA: m_2 = E_{PK_{CSPA}}\{K_{CSPA}, t_2\},\]
            where 
            \[K_{CSPA} = s \cdot H_1(ID_{CSPA}).\] 
        \end{enumerate} 
        
\subsection{EV Registration Phase}
        \begin{enumerate}
            \item EV generates its $ID_{EV}$ and sends the parameters to RA along with timestamp $t_3$:
            \[EV \rightarrow RA: m_3 = E_{PK_{RA}}\{ID_{EV}, t_3\}.\]
            \item RA chooses a seed $d_{EV}$ for each EV and a random number $a_i$ to generate pseudonyms and the ID-based private key for EV as:
            \[PID^i_{EV} = H_4(ID_{EV} || d_{EV} * a_i),\]
            \[K_{EV} = s \cdot H_1(ID_{EV}).\]
            \item RA transmits a message with the ID-based secret keys and other parameters to the EV:
            \[RA \rightarrow EV: m_4 = E_{PK_{EV}}\{K_{EV}, a_i, d_{EV}, \{k_i\}, t_4\},\]
            \[k_i = h(sH_1(ID_{RA}), H_2(ID_{RA} || PID^i_{EV})).\]
            \item EV extract the values $d_{EV}$ and $a_i$ and computes its own pseudonym values. 
            \item RA stores in a database the value $ID_{EV}$ along with the values of the seed and the random, and the $PID^{i}_{EV}$. From now on, RA and EV can communicate encrypting the messages with the ID-based public key cryptography.
        \end{enumerate}

\subsection{CSPA Authentication Phase}
        \begin{enumerate}
         \item EV sends its pseudonym to the bank in order to get a ticket for the billing process 
         \[EV \rightarrow Bank: m'_3 = E_{PK_{Bank}}\{PID^{i}_{EV}, t'_3\}.\]
         \item The Bank associates the $PID^{i}_{EV}$ to a newly generated ticket and sends it to EV
         \[Bank \rightarrow EV: m'_4 = E_{PK_{EV}}\{ticket, t'_4\}.\]
         This exchange with the Bank can be computed offline before using the ticket.  The Bank is considered a trusted party that knows the EV identity and the corresponding PID generation parameters to verify the correspondence between the PID and EV.
         \item EV generates a nonce $N_{EV}$ and a random number $r_{EV}$. It sends them to CSPA\\
        \[EV \rightarrow CSPA: m_5 = \{N_{EV}, r_{EV} \cdot P,  t_5\}.\] 
        \item CPSA generates a nonce $N_{CSPA}$ and a random number $r_{CSPA}$. It sends them to EV\\
        \[CSPA \rightarrow EV: m_6 = \{N_{CSPA}, r_{CSPA} \cdot P, ID_{CSPA}\]
        \[, t_6\}.\]
        \item EV uses the $PID^i_{EV}$ and computes
        \[k_{EV} = e(H_1(ID_{CSPA}), k_i),\]
        \[k'_{EV} = r_{EV} \; r_{CSPA} \;P,\]
        \[mac_{EV} = H_3(k_{EV} || k'_{EV} || ID_{CSPA} || PID^i_{EV} || N_{EV} \]
        \[|| N_{CSPA} || 00).\]
        EV transmits this $mac_{EV}$ value along with other parameters and sends it to the CSPA:
        \[EV \rightarrow CSPA: \] 
        \[m_7 = E_{ID_{CSPA}}\{ID_{RA}, PID^i_{EV}, ticket, mac_{EV}, t_7\}.\]
        \item CSPA decrypt the message and computes $mac^*_{EV}$ and compares it against the received $mac_{EV}$:
        \[k_{CSPA} = e(sH_1(ID_{CSPA}), h(H_1(ID_{RA}),\]
        \[H_2(ID_{RA} || PID^i_{EV}))),\]
        \[k'_{CSPA} = r_{EV} \; r_{CSPA} \;P,\]
        \[mac^*_{EV} = H_3(k_{CSPA} || k'_{CSPA} || ID_{CSPA} || PID^i_{EV}\]
        \[|| N_{EV} || N_{CSPA} || 00).\]
        \item CSPA generates a token $T$, computes $mac_{CSPA}$ and sends it to the EV along with a timestamp
        \[mac_{CSPA} = H_3(k_{CSPA} || k'_{CSPA} || ID_{CSPA} || PID^i_{EV} \]
        \[|| N_{EV} || N_{CSPA} || 10),\]
        \[CSPA \rightarrow EV: m_8 = \{mac_{CSPA}, T \oplus PID^i_{EV}, t_8\}.\]
        \item EV can compute $mac^*_{CSPA}$ and verify with the received value of $mac_{CSPA}$. 
        \[mac^*_{CSPA} = H_3(k_{EV} || k'_{EV} || ID_{CSPA} || PID^i_{EV}\]
        \[|| N_{EV} || N_{CSPA} || 10).\]
        \item If the verification is successful, it can extract $T$ from the message and store it for later use.  Both EV and CSPA can also compute autonomously a session key on the bi-linear map as:
         \[sk_{EV} = H_3(k_{EV} || k'_{EV} || ID_{CSPA} || PID^i_{EV} || N_{EV}\]
        \[|| N_{CSPA} || 01),\]
        \[sk_{CSPA} = H_3(k_{CSPA} || k'_{CSPA} || ID_{CSPA} || PID^i_{EV} \]
        \[|| N_{EV} || N_{CSPA} || 01).\]
        \item After authenticating the EV, CSPA computes the hash of the token $h(T)$ and encrypts it using the group key between RSU and CSPA, $GK_{RSU-CSPA}$ before sending it to all the RSUs\\
        \[CSPA \rightarrow RSU:\]
        \[ m_9 = E_{GK_{RSU-CSPA}}\{h(T),PID^i_{EV}, sk_{CSPA}, t_9\}.\]
        \end{enumerate} 
        
\subsection{RSU Authentication Phase}
        \begin{enumerate}
        \item After receiving a token, if the EV wants to initiate a charging session, it generates a nonce for the RSU, $N_{RSU}$ and transmits $m_{10}$ to the nearest RSU to initiate the charging process. \\
        \[EV \rightarrow RSU_j: m_{10} = E_{sk_{EV}}\{PID^i_{EV}, N_{RSU}, t_{10}\}.\]
        \item RSU receives $m_{10}$ and decrypts it to obtain the pseudonym of the EV. The RSU sends an acknowledgement $N_{RSU} + 1$ and a nonce $M_{EV}$ to the requesting EV listing the unit cost per charging pad $C_j$,\\
        \[RSU_j \rightarrow EV: \]
        \[m_{11} = E_{sk_{CSPA}}\{N_{RSU} + 1, M_{EV}, t_{11}, C_j\}.\]
        \item RSU then generates a hash chain and broadcasts it to all CPs belonging to it using the group key between the RSU and the CPs, $GK_{RSU_j - CP}$\\ 
        \[RSU_j \rightarrow CP:\]
        \[m_{12} = E_{GK_{RSU_j-CP}}\{h^n(h(T) || h(M_{EV})), \] 
        \[h^2(h(T)\oplus h(M'_{EV})), t_{12}\},\]
        \[\text{where }M'_{EV} = M_{EV} + k\]
        \end{enumerate} 
        
\subsection{CP Authentication Phase}
        \begin{enumerate}
        \item After receiving the token from the CSPA and the nonce from RSU, the EV computes the hash chain $h^n(h(T) || h(M_{EV}))$, sends $m_{13}$ to the CP,\\
        \[EV \rightarrow CP: m_{13} = \{1, h^{n-1}(h(T) || h(M_{EV}))\}.\]
        \item CP observes the first bit of the received message and if it is 1, the CP validates the hash, comparing (denoted by the $\stackrel{?}{=}$ operator) the hash of the value got by EV with the head of the hash chain.
        \[h(h^{n-1}(h(T) || h(M_{EV}))) \stackrel{?}{=} h^{n}(h(T) || h(M_{EV})).\] 
        \item If the authentication is successful, the charging plate will begin the transfer of electric power to the vehicle and transmits the received hash value from EV to the nearest neighbour CP, \\
        \[CP_i \rightarrow CP_{i+1}:\]
        \[m_{14} = E_{GK_{CP_j}}\{h^{n - (i + 1)}(h(T) \parallel h(M_{EV}))\}.\]
        \end{enumerate}
        
\subsection{Charging Termination Phase}
        \begin{enumerate}
        \item When the EV wants to terminate the charging process, it can compute $r'_{EV} = r_{EV} + k$ (k constant published by RA) and send $m_{14}$ to the upcoming CP,
        \[EV \rightarrow CP_i: m_{15} = \{0, h(h(T)\oplus h(M'_{EV})), \hat{y}\},\]
        \[ \text{where } \hat{y} = h^2(h(T)\oplus h(M'_{EV})).\]
        \item CP observes the first bit of the received message and if it is 0, CP can compare the subsequent terms of the message and verify that they are the same,
        \[h(h(h(T)\oplus h(M'_{EV}))) \stackrel{?}{=} \hat{y}.\] 
        \item If the verification is successful, the CP can send the hash value it received from its neighbour to the RSU,
        \[CP_i \rightarrow RSU_j: m_{16} = E_{GK_{RSU_j-CP}}\{\hat{y}, t_{16}\},\] 
        and terminate the charging cycle for the EV.
        \end{enumerate}

    \section{Proposed Billing Protocol}

\begin{enumerate}
    \item RSU can decrypt $m_{13}$ and obtain the value of $PID^{i}_{EV}$ and $h(T)$ directly looking in the database against the value of $\hat{y}$ received.
    \item RSU can compute the cost as $Cost_{RSU_j} = l * C_j$ (with $l$ number of pads crossed) and send $m_{17}$ to the CSPA,
    \[RSU_j \rightarrow CSPA:\]
    \[ m_{17} = E_{GK_{RSU-CSPA}}\{h(T),Cost_{RSU_j},t_{17}\}.\]
    \item When the CSPA receives $m_{17}$ from the RSU, it can identify the pseudonym of the vehicle based on the received hash value of token.
    \item CSPA will make an entry for the pseudonym of the vehicle with the respective cost received. 
    \item If the EV restarts the charging cycle before the expiry of the token, the sum of all costs for each RSU gets accumulated against the pseudonym.
    \[Cost_{PID^i_{EV}} = \sum_{j=1}^{m} Cost_{RSU_j},\]
    where $m$ represents the charging cycle.    
    \item When the token expires, the CSPA bills the bank for the total cost accumulated against the EV.
    \[CSPA \rightarrow Bank:\]
    \[ m_{18} = E_{PK_{Bank}}\{ticket,Cost_{PID^{i}_{EV}}, t_{18}\}.\]
    
    \item Bank pays the CSPA and bills the vehicle accordingly, checking the payment information against the ticket:
    \[Bank \rightarrow EV: m_{19} = E_{PK_{EV}}\{Cost_{PID^{i}_{EV}}, t_{19}\}.\]
    \item EV checks the bill with its own cost estimation and if it does not match the value measured by the OBU, he/she can decide to dispute the billing. The Tamper-Proof OBU is a deterrent of disputing every bill, due to the possibility to check immediately the veracity of the driver argument about the unfair billing.
    
\end{enumerate}

    \section{Formal Security Analysis}
We provide a formal security analysis in Sections VII.A and VII.B using Random Oracle Model and BAN logic, respectively. Furthermore, we describe an informal security analysis using STRIDE model in Section VII.C.

\subsection{Security analysis using ROR model}
In this subsection, we discuss the security of the protocol in terms of impracticability for the attacker to infer the session key computed between the EV and the CSPA. In order to do so, we rely on the well-known \textit{Real-Or-Random} (\textit{rr} or \textit{ROR}) model, as first discussed in \cite{Bellare} and then extended in \cite{Abdalla}. The model, as described by Abdalla \textit{et al.}, utilizes different Random Oracles, as follows. At the beginning of the experiment, a bit \textit{b} is chosen randomly and uniformly in ${0,1}$, then we have the following oracle functions.

\begin{itemize}
    \item \textit{Execute}: model a passive attack as an eavesdropper between the client and the server. The query returns the messages exchanged;
    \item \textit{Send}: model an active attack as an adversary that can intercept, forge, and modify the messages. The query returns the message generated by the receiver after getting the adversary's packet;
    \item \textit{Test}: returns the session key if $b = 1$ or a random key if $b = 0$.
\end{itemize}

An adversary $\mathcal{A}$ can make any queries to \textit{Test} as she/he wants, but depending on the initial bit \textit{b},  it will always reveal a real or a random key. If the key returned by the oracle is random, this must be the same value for all the related instances of the experiment. \par
The goal of the adversary $\mathcal{A}$ is to guess the hidden bit \textit{b} correctly and therefore, the success is measured in terms of indistinguishability for the adversary to determine if a given key by the \textit{Test} oracle is real or random. According to \cite{Abdalla}, the advantage of adversary $\mathcal{A}$ in this setting is:

\[Adv^{ror}_{P, D}(\mathcal{A}) = | 2\cdot Pr[Succ]-1 |,\]

where \textit{P} is the protocol instance and \textit{D} the session key space. \par
Now, we can  model two possible games for the adversary $\mathcal{A}$.

\begin{itemize}
    \item $Game_0$: the adversary is passive; therefore she/he cannot learn anything about the session key from intercepted messages. In this game,  $\mathcal{A}$ can only try to guess if the output of the oracle is random or not. The advantage for the adversary in this game is $Adv^{ror}_{P, D}(\mathcal{A}) = | 2\cdot Adv_{Game_0}(\mathcal{A})-1 |$, having to guess \textit{b} without information.
    \item $Game_1$: in this case, the adversary has access as an active participant to the hash function $H_3$ and the Oracle itself (ideal hash version). The goal, as before, is to identify the hidden bit \textit{b}. $\mathcal{A}$ can now try to get the output of the function $H_3$ and then ask the Oracle to elaborate the same input. In terms of indistinguishability between the two games, we have that the probability for an adversary to get the correct value is bounded by the well-known result of the \textit{Birthday paradox}:
    \[|Adv_{Game_0}(\mathcal{A}) - Adv_{Game_1}(\mathcal{A})| \leq \frac{Q}{|\mathcal{H}|},\]
    with $Q$ number of queries to the hash function and $\mathcal{H}$ space of the hash function. The goal of $\mathcal{A}$ is still to guess \textit{b}, and after $Game1$, having a collision as the adversary desires, is $Adv_{Game_1}(\mathcal{A}) = 1/2$.
\end{itemize}

As final result, we have:

\[Adv^{ror}_{P, D}(\mathcal{A}) = | 2\cdot Adv_{Game_0}(\mathcal{A})-1 |,\]
\[\frac{1}{2} Adv^{ror}_{P, D}(\mathcal{A}) = | Adv_{Game_0}(\mathcal{A})-\frac{1}{2} |.\]

From $Game_1$:

\begin{gather*}
    \frac{1}{2} Adv^{ror}_{P, D}(\mathcal{A}) = | Adv_{Game_0}(\mathcal{A})-\frac{1}{2} |,  \\
    \frac{1}{2} Adv^{ror}_{P, D}(\mathcal{A}) \leq \frac{Q}{|\mathcal{H}|}, \\
    Adv^{ror}_{P, D}(\mathcal{A}) \leq \frac{2Q}{|\mathcal{H}|}.
\end{gather*}

This is the final result for the amount of adversary advantage, croncretely showing that for a \textit{collision-resistant}, \textit{one-way} hash function, the adversary cannot do better than guess the possible session key.

    \begin{table}[b!]
\renewcommand{\arraystretch}{1.5}
\begin{center}
    \begin{tabular}{| p{2cm}  p{5cm} |}
        \hline
        \textbf{NOTATION} & \textbf{DESCRIPTION} \\
        \hline
        $P |= X$ & P believes X, so P thinks that X is true. \\
        \hline
        $P <| X$ & P sees message X. \\
        \hline
        $\{X\}K$ & encrypted with key K.  \\
        \hline
        $P |\sim X$ & P once said X. \\
        \hline
        $\#(X)$ & X is fresh. \\
        \hline
        $P <- K -> Q$ & P and Q shared a secret key. \\
        \hline
        $P=X=Q$ & X is a secret known only by P and Q (or other trusted parties) \\
        \hline
        Shared Key Rule & $\frac{P |= Q <- K -> P, P <| \{x\}K}{P |= Q |\sim X}$, If P believes that K is a good K, and P sees X encrypted with K, then P believes that Q once said X.  \\
        \hline
        Nonce Verification Rule & $\frac{P |= \#(X), P |= Q |\sim X}{P |= Q |= X}$, the only formula in order to promote $|\sim$ to $|=$, says that P believes X to be recent, and Q said X, then P believes that Q believes X. \\
        \hline
        Freshness Rule & $\frac{P |= \#(X)}{P |= \#(X, Y)}$, if part of the formula is fresh, the entire formula is believed to be fresh. \\
        \hline
    \end{tabular}
    \caption{BAN Constructs}
    \label{table:ban_constructs}
\end{center}
\end{table}
    
    \subsection{Replay attack resistance with BAN Logic} 
This section is dedicated to the formal proof of authentication and replay attack resistance by means of BAN logic construction, first introduced by Burrows \textit{et al.} \cite{ban}. 

\subsubsection{Introduction to the BAN logic}
This approach aims to make explicit and clear assumptions on what the protocol needs to achieve and then describe it by formal algebra. In this perspective, an explicit initial state is required to derive relationships between state transactions in terms of \textit{belief}. At the end of the protocol, the final state will help make considerations about possible improvements. We use the following main principles,

\begin{itemize}
    \item Freshness: if a message with a nonce (or random number) is sent for the first time and a response based on that nonce is received, the later message is considered fresher than the first one.
    \item Shared secret: if two parties trust each other in sharing a valid secret, a message encrypting that secret is assumed to have been sent by one of the two intended users.
\end{itemize}

We describe the messages as $A \rightarrow B: \{X\}K_{ab}$, where the previous one is a message encrypted with a session key from A to B. A and B are the parties of the system, and they only state what they believe. When a party trusts something from the current state of the protocol, then we can say that it trusts all the other running instances. \par
All the elements have the same role and meaning as in the protocol described in the previous chapter, except for the use of the conjunction "\textit{,}" as a connector between propositions. The BAN logic is based on constructs that we report in Table \ref{table:ban_constructs}. 

\subsubsection{Logical proof for our protocol}
Considering the registration parts as secure, we can start the analysis with the first message exchanged between EV and CSPA, $m_5$.
Having a timestamp $t_x$ in all the messages, we can use the Freshness rule in order to provide security against a Replay attack for all the exchanges in the protocol. As a result, CSPA, after receiving $m_5$, considers fresh all the parameters after validating the timestamp:
\[CSPA |= \#(N_{EV}, r_{EV} \cdot P, t_{5}).\] 
The same applyes for $m_6$:
\[EV |= \#(N_{CPSA}, r_{CSPA} \cdot P, t_{6}).\] 

The most important part of the protocol is in the next two messages. In $m_7$ EV and CSPA share parameters that are secret to all the other parties (e.g. an adversary).
The freshness of the $PID^i_{EV}$ can be checked also against the database that CSPA builds using the previous instances of the protocol. Assuming the two parties are legitimate, CSPA can check the value of $mac_{EV}$ (as the protocol expect) and in case they match we have:

\[ \frac{CSPA|= EV = mac_{EV} = CSPA}{CSPA |= EV |\sim \{PID^i_{EV}, mac_{EV}, t_7\}} ,\]
\[\frac{CSPA |= EV = PID^i_{EV} = CSPA}{CSPA |= EV |\sim \{PID^i_{EV}, mac_{EV}, t_7\}} , \]
\[\frac{CSPA <| \{PID^i_{EV}, mac_{EV}, t_7\}K_{ID_{CSPA}}}{CSPA |= EV |\sim \{PID^i_{EV}, mac_{EV}, t_7\}}.\]
 
 Due to the freshness of the parameters and verified $mac_{EV}$, we can state that CSPA believes EV in the parameters that it sends and can compute the session key $SK_{CSPA}$:
 \begin{gather*}
    CSPA |= EV |\sim \{mac_{EV}\}, \\
    CSPA |= EV |= \{mac_{EV}\}, \\
    CSPA |= EV <- SK_{CSPA} -> CSPA.
\end{gather*}

The same applies for the next message $m_8$. EV can check the value of $mac_{CSPA}$ against its own computation, then if it is valid we can state:
\[\frac{EV <| \{mac_{CSPA}, t_8)\}}{EV |= CSPA |\sim \{mac_{CSPA}, t_8)\}}.\]
Resulting in:
 \begin{gather*}
    EV |= CSPA |\sim \{mac_{CSPA}\}, \\
    EV |= CSPA |= \{mac_{CSPA}\}, \\
    EV |= CSPA <- SK_{EV} -> EV.
\end{gather*}
And in conclusion, for EV and CSPA authentication and Session Key derivation:
 \begin{gather*}
     CSPA |= EV |= EV <- SK_{CSPA} -> CSPA, \\
     EV |= CSPA |= CSPA <- SK_{EV} -> EV.
\end{gather*}
As we showed in the previous section, the Session Key computed by the two parties follow in the same value thank to the bilinear mapping and Weil Diffie-Hellman properties. \par
In addition to prove resistance against Replay attack, the last two messages show that even a modification or forging attack is not possible. The clear messages could be sent with a timestamp replacement in order to use the same values for the nonce and random point, but the $PID^i_{EV}$, as well as the $mac_{EV}$, remain secret to an adversary not knowing the private key of CSPA. Consequently, an adversary could play with $m_5$ and $m_6$ but the values for the $mac$ computed by the intended entities would be different, resulting in the impossibility to authenticate for the malicious user. \par

\begin{table*}[b!]
\begin{center}
    \begin{tabular}{| p{3cm} | p{3cm} | p{10cm} | }
        \hline
        \textbf{THREAT} & \textbf{PROPERTY}  & \textbf{DESCRIPTION}\\
        \hline
        Spoofing & Authenticity & Allows the attacker to pretend to be an authenticated user/trusted authority. \\
        \hline
        Tampering & Integrity & Attackers are able to make changes or modifications to stored/transmitted data without authorization. \\
        \hline
        Repudiation & Non-repudiability & Intruders are able to deny their denies their involvement with an attack and blame others. The system's capability to trace down malicious activity back to the attacker is limited.\\
        \hline
        Information Disclosure & Confidentiality & The system unintentionally reveals messages with sensitive information to unauthorized users.\\
        \hline
        Denial of Service & Availability & These attacks render an entity temporarily unusable and prevent a legitimate user from accessing those entities. \\
        \hline
        Elevation of Privilege & Authorization & A user, either authorized or unauthorized, can get access to information that they are not authorized to see.\\
        \hline
    \end{tabular}
    \caption{Threats in STRIDE Analysis}
    \label{table:stride}
\end{center}
\end{table*}

Once the authentication between EV and CSPA is concluded, the next step is the authentication between EV and RSU using the previusly computed Session Key. Always referencing the Freshness rule, we have that $m_{10}$ is considered new and legitimate by RSU, with logic expression as follow:
\[ \frac{RSU|= RSU <- SK -> EV}{RSU |= EV |\sim \{N_{RSU}, PID^i_{EV}, t_{10}\}} ,\]
\[\frac{RSU <| \{N_{RSU}, PID^i_{EV}, t_{10}\}}{RSU |= EV |\sim \{N_{RSU}, PID^i_{EV}, t_{10}\}},\]

and by Nonce Verification rule:
\begin{gather*}
     RSU |= EV |= \#(N_{RSU}),\\
     RSU |= \#(N_{RSU}).
\end{gather*}
In message $m_{11}$, EV can check whether the Nonce value has been updated and encrypted with the Session Key, meaning that RSU could decrypt it correctly and answer back:
\[ \frac{EV|= EV <- SK -> RSU}{EV |= RSU |\sim \{N_{RSU} + 1, N_{EV}, t_{11}\}} ,\]
\[\frac{EV <| \{N_{RSU} + 1, N_{EV}, t_{11}\}}{RSU |= EV |\sim \{N_{RSU} + 1, N_{EV}, t_{11}\}}.\]

As before, from Nonce Verification rule and the verified updated value of $N_{RSU} + 1$, EV trust RSU, leading to the acceptance of the Nonce used in the hash-chain:
\begin{gather*}
     EV |= RSU |= \#(N_{EV}),\\
     EV |= \#(N_{EV}).
\end{gather*}

This concludes the proof, showing the validity of the session key and the impossibility to perform a replay attack. Furthermore, we showed that even a modification or forging attack (as well as a Man-In-The-Middle) is impractical for an adversary in that it results in an authentication denial.
    
    \section{STRIDE Analysis}

Threat modeling is a process using which we could identify threats, attacks and vulnerabilities that could impact our system. In this section, we use STRIDE, a threat modeling framework developed by Microsoft, to evaluate the security of our proposed authentication protocol. STRIDE stands for Spoofing identity, Tampering with data, Repudiation threats, Information disclosure, Denial of service and Elevation of privileges\cite{stride}. Each one of the above threats can be matched to a desired property, which along with a description about the threat, is summarised in Table \ref{table:stride}. This section deconstructs the authentication protocol and applies a STRIDE analysis to each of them. 

\subsection{Spoofing} 

For a spoofing attack, the adversary would have to pretend to be the EV user and communicate with other entities in the architecture. In this protocol, the adversary cannot impersonate an EV because, during authentication, an adversary does not know the pseudonym that the intended EV would use. This is because the pseudonym is unique every time and cannot be generated by entities other than the intended EV itself.

\par The possible messages the adversary might try to spoof are $m_3, m_7, m_{10} \text{ and } m_{13}$. With $m_3$, he/she can use a random ID and pretend to be the user. However, the RA will not authenticate spoofed vehicles. By generating a random pseudonym, the adversary can try to spoof $m_7$. However, the CSPA will not issue the token T without verifying $mac_{EV}$, which is prevented with the help of bilinear mapping. In the case of $m_{10}$, the RA would reject the message if it was encrypted with any key other than $sk_{EV}$, which the adversary does not have access to. The adversary requires the value of token T to spoof the hash-chain sent to the CP in $m_{13}$, which it cannot obtain when $m_7$ is discarded.

\subsection{Tampering}

The adversary may tamper with the EV's messages before sending them to the CSPA, RSU, or CP. The adversary may decide to tamper with the messages transmitted by the EV before they are sent to CSPA, RSU, or CP. The messages $m_3, m_7, m_{10}$ are encrypted with the public key of RA, ID of the CSPA, and the session key generated between EV and CSPA, respectively. Because the adversary cannot decrypt any of these messages, it is unable to tamper with them.


\subsection{Repudiation}

Introducing unique pseudonyms to each registered EV reduces the opportunity for an EV to deny the messages it transmits. The messages sent by the EV always have either the identity of the EV ($ID_{EV}$) or the pseudonym ($PID^i_{EV}$), which can be traced back to the EV's identity. Message $m_{13}$, which does not have either of the above components, has a token T which is unique for each EV and known to the CSPA. Thus, it can be traced back to the corresponding EV as well. Hence, an adversary pretending to be an EV cannot repudiate any of the messages sent.

\subsection{Information Disclosure}

Most message exchanges in the protocol are encrypted with appropriate encryption keys to prevent sensitive information from being leaked - $m_1, m_2$ encrypted with public keys of RA and CSPA, respectively, $m_3, m_4$ encrypted with public keys of RA and EV, respectively, and $m_7$ encrypted with CSPA's ID. To encrypt $m_9, m_{12}$, we use the group keys between RSU - CSPA and the group key between RSU - CP.
The messages sent in the open are $m_5, m_6, m_{15}$, which contain nonces and timestamps. As these exchanges do not contain any sensitive information, any adversary who obtains them will be unable to decipher any useful information. In $m_8$ as well, the $mac_{EV}$ is an hashed value and the other parameter is the result of an XOR operation, neither of which would prove beneficial for the adversary.

\subsection{Elevation of Privilege}

As discussed earlier, for information disclosure, message exchanges between any two parties in the protocol are encrypted with keys decipherable only by the entity for which the message is intended. Even if unauthorized parties were to get access to the transmitted messages used for derive a session key between EV and CSPA, they will not be able to use any of that information for their benefit thanks to the use of bilinear mapping in the Identity-Based Public Key encryption. 
    
    \section{Performance Analysis}

In this section, we consider the computation and communication costs incurred by the protocol during the authentication process. In the first part of the section, we describe the experiment setup and execution while in the second part, we compare the performance of our work against that of other recent and SOTA protocol schemes. In the last section, we show the communication cost of our implementation.

\subsection{Experimental Analysis}
We simulate the performance of the protocol using the Charm-Crypto Python Library \cite{charm}. This library provides methods for simulating and testing the cryptographic protocols, embedding a benchmark tool. The experiment is performed on a laptop with the following specifics: 8 GB of memory, 256 GB of SSD space, Intel Core i7-6500U @ 2.50Ghz x 4, Ubuntu 20.04.3 LTS with 64-bit architecture. The library allows the calculation of the computation time for the three primitives used in our protocol, namely: multiplication, exponentiation and pair in a bilinear map. The simulation of these primitives is executed 1000 times to get the average, the min and the max values with high confidence. The values $T_{mul}$, $T_{exp}$ and $T_{pair}$, respectively the time taken for multiplication, exponentiation and map, are listed in Table \ref{table:computational_cost}. $T_h$ is the time taken by a single SHA-256 hash operation. The other values reported in the Table are the time for the operations used in \cite{elghanam} ($T_E$, $T_S$, $T_{dE}$, $T_V$, corresponding to the time for encryption/decryption and for signing/verification in RSA) and in \cite{babu} ($T_{ecm}$, $T_{sig}$, $T_{ver}$: point multiplication in Elliptic curves, time for signing and verification in ECDSA). 

\subsection{Computational cost}
In Table \ref{table:comparison_cost}, we report the comparison of the computational cost of our authentication protocol to authentication protocols with a similar architecture. Our scheme has a computation overhead slightly higher than the most recent and lightweight protocol of Babu \textit{et al.} \cite{babu}.

\begin{table}[t!]
\begin{center}
    \begin{tabular}{| p{1.5cm} | p{1.5cm} | p{1.5cm} | p{1.5cm}|}
        \hline
        \textbf{Primitive} & \textbf{Average Time (ms)}  & \textbf{Min. Time (ms)} & \textbf{Max. Time (ms)}\\
        \hline
        $T_{mul}$ & 0.005 & 0.004 & 0.005 \\
        \hline
        $T_{exp}$ & 0.110 & 0.102 & 0.630 \\
        \hline
        $T_{pair}$ & 0.884 & 2.763 & 0.833 \\
        \hline
        $T_{h}$ & 0.27 & 0.266 & 0.594 \\
        \hline
        $T_{ecm}$ & 1.352 & 1.352 & 1.352 \\
        \hline
        $T_{ver}$  & 1.449 & 1.449 & 1.449 \\
        \hline
        $T_{sig}$  & 0.992 & 0.992 & 0.992 \\
        \hline
        $T_{E}$ & 3.925 & 3.749 & 7.31  \\
        \hline
        $T_{dE}$ & 4.047 & 3.801 & 7.53  \\
        \hline
        $T_{S}$ & 4.046 & 3.918 & 7.767  \\
        \hline
        $T_{V}$ & 4.008 & 3.883  & 7.436  \\
        \hline
    \end{tabular}
    \caption{Cryptographic primitive execution time}
    \label{table:computational_cost}
\end{center}
\end{table}

\begin{table*}[!ht]
\begin{center}
    \begin{tabular}{| p{1.5cm} | p{2.7cm} | p{2.7cm} | p{2.7cm} | p{2.7cm} | p{2.7cm} |}
        \hline
        \textbf{Entity} & \textbf{Roman \textit{et al.}  \cite{roman}} & \textbf{ElGhanam \textit{et al.} \cite{elghanam}} & \textbf{Babu \textit{et al.} \cite{babu}} & \textbf{Proposed (without on-demand charging)} & \textbf{Proposed (with on-demand charging)}\\
        \hline
        $EV$ & $3T_{ecm}+2T_{pair}+T_{ver}+5T_h\approx7.968$ ms & $3T_E+3T_{dE}+3T_S+3T_V\approx48.078$ ms & $T_{ecm}+4T_{h}\approx2.432$  ms & $3T_{mul}+1T_{exp}+2T_{pair}+5T_{h}\approx3.243$ ms & $3T_{mul}+1T_{exp}+2T_{pair}+7T_{h}\approx3.783$ ms\\
        \hline
        $CSPA / FS$ & $T_{ecm}+2T_{pair}+T_{sig}+2T_h+\beta T_h\approx0.27\beta+4.652$ ms & $2T_E+2T_{dE}+2T_S+2T_V\approx$32.052 ms & $2T_{ecm}+2T_{h}\approx3.244$ ms & $3T_{mul}+T_{pair}+5T_{h}\approx2.249$ ms & $3T_{mul}+T_{pair}+6T_{h}\approx2.519$ ms\\
        \hline
        $RSU$ & $5T_h\approx1.35$ ms & $T_E+T_{dE}+T_S+T_V\approx16.026$ ms & $6T_{h}\approx1.62$ ms & $2T_{h}\approx0.54$ ms & $4T_{h}\approx1.08$ ms\\
        \hline
        $CP$ & $((n^2+n)/2)T_h$ & $((n^2+n)/2)T_h$ & $((n^2+n)/2)T_h$  & $((n^2+n)/2)T_h$ & $((n^2+n)/2)T_h$\\
        \hline
        $TOT$ & $13.97+ \beta 0.27+T_{CP}$ ms & $ 96.156+T_{CP}$ ms & $5.676+T_{CP}$ ms & $6.062+T_{CP}$ ms & $7.382+T_{CP}$ ms\\
        \hline
    \end{tabular}
    \caption{Comparison of Computational Costs}
    \label{table:comparison_cost}
\end{center}
\end{table*}

In our case, we can introduce a public key encryption system without losing performance compared with a system composed of hash and XOR operations. Furthermore, the other schemes significantly differ in the time they take for computation. 

The hash-chain approach is used between EV and CP in all the protocols as it is fast and lightweight. The computational overhead derived from the authentication with the pads is a function of the number of pads used, so the time $T_{CP}$ refers to the total time of the authentication and charging process in the lane. The time for the hash computation is the same for all the schemes compared. The symmetric key operations are considered negligible in all the cases.

\subsection{Communication cost}

There are only four messages exchanged between EV and CSPA in the first part of the authentication in our protocol. Almost all the parameters are generated, and the authentication requires the most expensive messages and computations. After the EV and CSPA authenticate each other, only two messages between EV and RSU are necessary. It consists only of a challenge to confirm that the RSU can decrypt the nonce correctly, meaning that it is the intended RSU that received the session key by CSPA. The remaining part of the protocol is almost the same for all the proposed schemes in the last years, with the implementation of a hash chain to verify the EV with the pads. The messages are lightweight and fast, considering they are composed only of a hash (fixed length) and a byte prepended to communicate if the EV wants to continue the charging process or terminate it. 

\begin{table*}[t!]
\begin{center}
    \begin{tabular}{| p{1cm} | p{6cm} | p{3cm} |}
        \hline
        \textbf{Message} & \textbf{Content}  & \textbf{Message length (B)}\\
        \hline
        $m_5$ & $\{N_{EN}, r_{EV} \cdot P, t_5\}$ & 32 + 64 + 8 = 104 B  \\
        \hline
        $m_6$ &  $\{N_{CSPA}, r_{CSPA} \cdot P, ID_{CSPA}, t_6\}$ & 32 + 64 + 32 + 8 = 136 B   \\
        \hline
        $m_7$ & $\{ID_{RA}, PID^i_{EV}, ticket, mac_{EV}, t_7\}$ & 32 + 32 + 64 + 32 + 8 = 170 B \\
        \hline
        $m_8$ &  $\{ mac_{CSPA}, T \oplus PID^i_{EV}, t_8\}$ & 32 + 32 + 8 = 72 B \\
        \hline
        $m_9$ & $\{ h(T), PID^i_{EV}, sk_{CSPA}, t_9\}$ & 32 + 32 + 32 + 8 = 104 B \\
        \hline
        $m_{10}$  & $\{ PID^i_{EV}, N_{RSU}, t_{10}\}$ & 32 + 32 + 8 = 72 B \\
        \hline
        $m_{11}$  & $\{N_{RSU}+1, N_{EV}, t_{11}, C_j\}$ & 32 + 32 + 8 + 1 = 73 B  \\
        \hline
        $m_{12}$ & $\{ h^n(h(T) || h(N_{EV})), h^2(h(T)\oplus h(N'_{EV})), t_{12}\}$ & 32 + 32 + 8 = 72 B    \\
        \hline
        $m_{13}$ & $\{1, h^{n-1}(h(T) || h(N_{EV}))\}$ & 1 + 32 = 33 B \\
        \hline
        $m_{14}$ & $\{h^{n - (i + 1)}(h(T) || h(N_{EV}))\}$ & 32 B   \\
        \hline
        $m_{15}$ &  $\{0, h(h(T)\oplus h(N'_{EV})), \hat{y}\}$ & 1 + 32 + 32 = 65 B \\
        \hline
        $m_{16}$ & $\{\hat{y}, t_{16}\}$ & 32 + 8 = 40 B \\
        \hline
    \end{tabular}
    \caption{Communication cost of each message during authentication.}
    \label{table:communication_cost}
\end{center}
\end{table*}

\par In Table \ref{table:communication_cost} the communication cost associated with each message is computed. We used a 32-byte ID for the entities, a 32-byte pseudonym, an 8-byte timestamp, hash functions with a digest of 32-byte, along with 16-byte seed and update parameters for pseudonyms.

    \section{Discussion}
The advancement in dynamic wireless charging necessitates a secure authentication and authorization scheme in a short time. Although the technology is in an early stage of development, the researchers and companies need to be prepared with a secure infrastructure when the deployment for the mass comes. In this way, the proposed scheme considers the general representation of what could be the actual scenario and the requirements of fast computation and privacy-preserving. The physical construction of the roads and the pads constitute the most challenging part, from which the authentication scheme could depend. For example, the length of the pad is a critical factor in the security of the system, making the defense against free-riding a challenging problem in the case of a very long plate. In our work, we considered short pads, each one spaced from the others. The necessity of a fast computation is the foundation of the research that has been done in that it is a requirement that cannot be neglected.
    
    \section{Conclusion}
In this paper, we proposed a new authentication and billing protocol scheme based on ID public-key cryptography and the Diffie-Hellman problem. The protocol allows the customer to start and stop the charging process without deciding in advance the amount of energy that she/he wants to purchase. To the best of our knowledge, this strategy has not been accounted for in the existing protocols. We carefully designed the protocol to have a fast implementation to meet the constrained system's requirements. This is suitable for implementing dynamic wireless charging which needs security and safety for the customer.

    \begin{IEEEbiography}[{\includegraphics[width=1in,height=1.25in,clip,keepaspectratio]{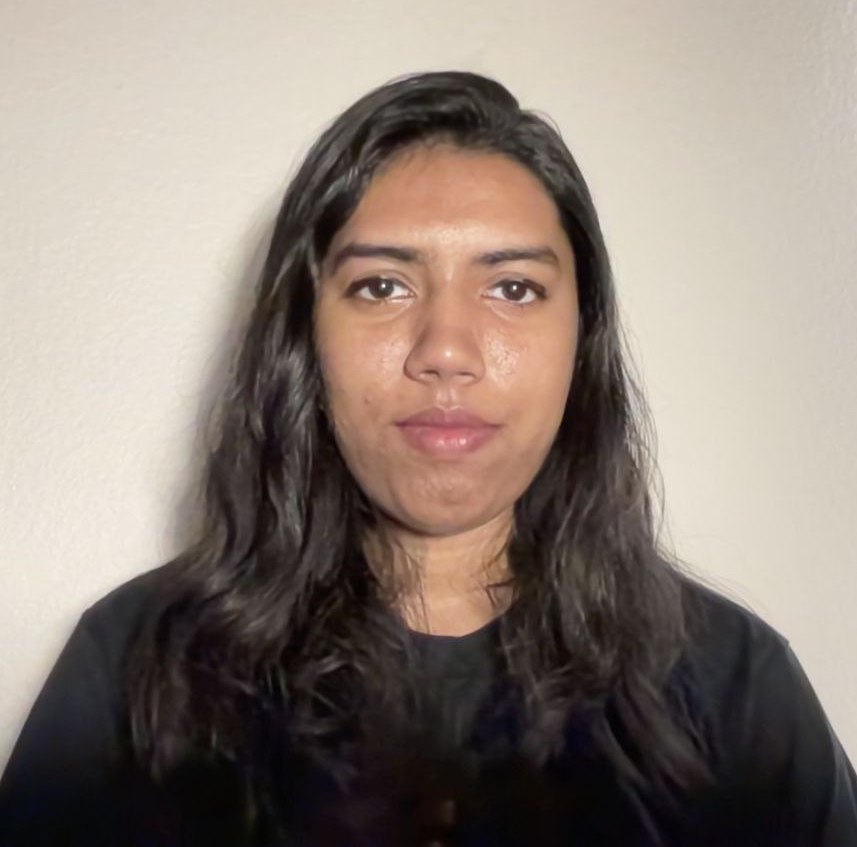}}]{Surudhi Asokraj}
    received her Bachelors in Electronics and Communications Engineering from SSN College of Engineering, India, in 2020. She is currently working towards her Master's degree in Electrical engineering in the Department of Electrical and Computer Engineering at the University of Washington, Seattle, with Network Security as her concentration. Her research interests include the security and privacy of electric vehicles, cyber-physical systems, and networking.
    \end{IEEEbiography}
    
    \begin{IEEEbiography}[{\includegraphics[width=1in,height=1.25in,clip,keepaspectratio]{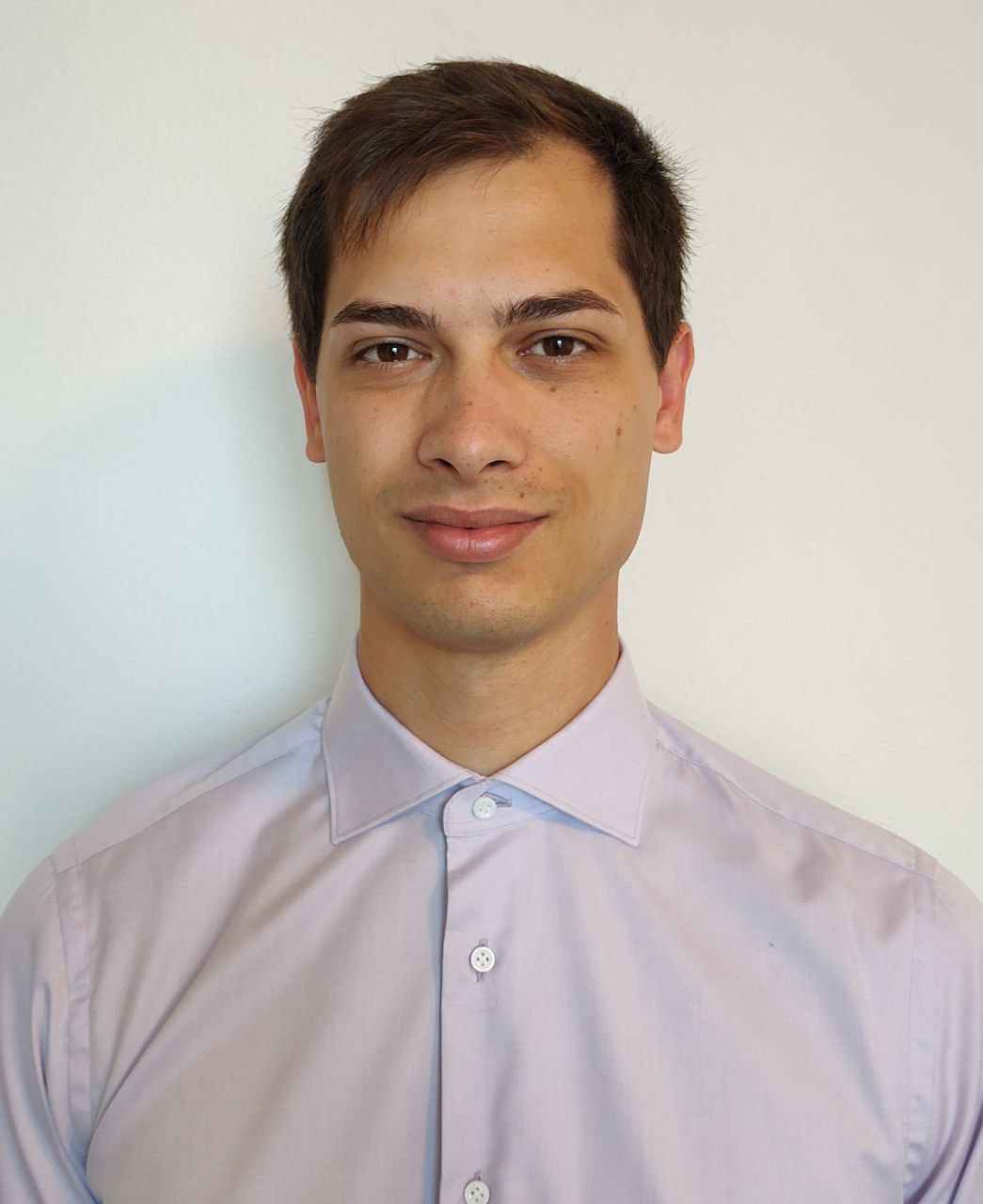}}]{Tommaso Bianchi} is a MSc student in ICT for Internet and Multimedia at the University of Padova, Italy. He is currently a visiting student researcher at the University of Washington, Seattle (WA). He received his Bachelors in Computer Engineering at the the University of Padua. His research interests include cyber-physical systems, vehicular security, software and mobile security, and binary analysis.
    \end{IEEEbiography}
    
    \begin{IEEEbiography}[{\includegraphics[width=1in,height=1.25in,clip,keepaspectratio]{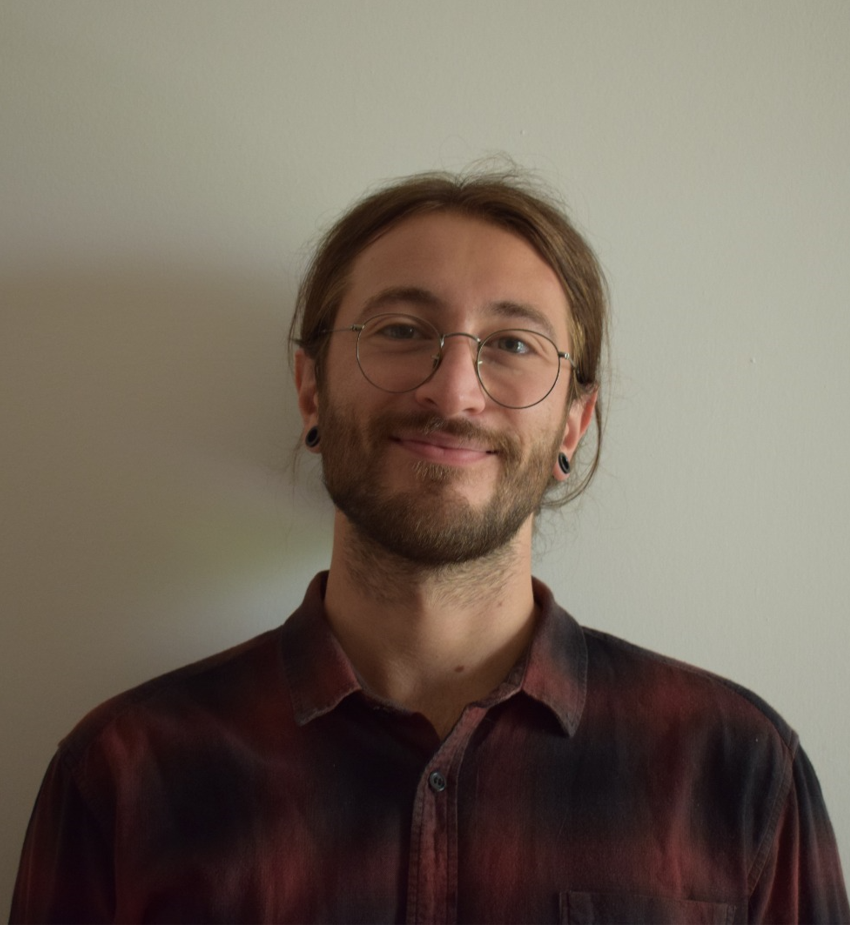}}]{Alessandro Brighente} is a postdoctoral researcher at the University of Padova. He received his Ph.D. degree in Information Engineering from the University of Padova in Feb. 2021. From June to November 2019 he was visitor researcher at Nokia Bell Labs, Stuttgart. He has been involved in European projects and company projects with the University of Padova. He served as TPC for several conferences, including Globecom and VTC. He is guest editor for IEEE Transactions on Industrial Informatics. His current research interests include security and privacy in cyber-physical systems, vehicular networks, blockchain, and physical layer security.
    \end{IEEEbiography}
    
    \begin{IEEEbiography}[{\includegraphics[width=1in,height=1.50in,clip,keepaspectratio]{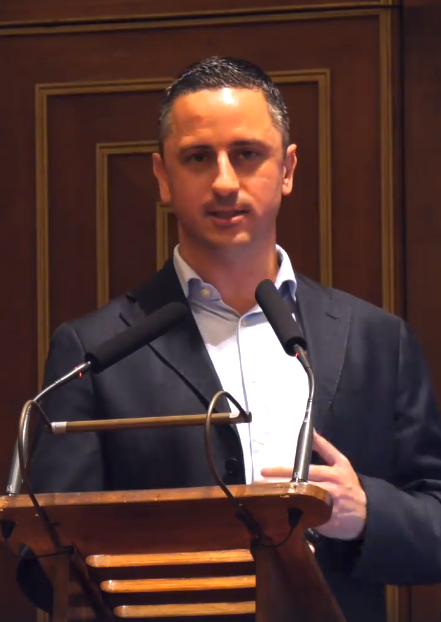}}]{Mauro Conti} is Full Professor at the University of Padua, Italy. He is also affiliated with TU Delft and University of Washington, Seattle. He obtained his Ph.D. from Sapienza University of Rome, Italy, in 2009. After his Ph.D., he was a Post-Doc Researcher at Vrije Universiteit Amsterdam, The Netherlands. In 2011 he joined as Assistant Professor at the University of Padua, where he became Associate Professor in 2015, and Full Professor in 2018. He has been Visiting Researcher at GMU, UCLA, UCI, TU Darmstadt, UF, and FIU. He has been awarded with a Marie Curie Fellowship (2012) by the European Commission, and with a Fellowship by the German DAAD (2013). His research is also funded by companies, including Cisco, Intel, and Huawei. His main research interest is in the area of Security and Privacy. In this area, he published more than 400 papers in topmost international peer-reviewed journals and conferences. He is Editor-in-Chief for IEEE Transactions on Information Forensics and Security, Area Editor-in-Chief for IEEE Communications Surveys \& Tutorials, and has been Associate Editor for several journals, including IEEE Communications Surveys \& Tutorials, IEEE Transactions on Dependable and Secure Computing, IEEE Transactions on Information Forensics and Security, and IEEE Transactions on Network and Service Management. He was Program Chair for TRUST 2015, ICISS 2016, WiSec 2017, ACNS 2020, CANS 2021, and General Chair for SecureComm 2012, SACMAT 2013, NSS 2021 and ACNS 2022. He is Fellow of the IEEE, Senior Member of the ACM, and Fellow of the Young Academy of Europe.
    \end{IEEEbiography} 
    
    \begin{IEEEbiography}[{\includegraphics[width=1in,height=1.25in,clip,keepaspectratio]{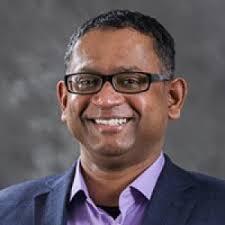}}]{Radha Poovendran}(F'15) is a Professor in the Department of Electrical and Computer Engineering at the University of Washington (UW) - Seattle. He served as the Chair of the Electrical and Computer Engineering Department at UW for five years starting January 2015. He is the Director of the Network Security Lab (NSL) at UW. He is the Associate Director of Research of the UW Center for Excellence in Information Assurance Research and Education. He received the B.S. degree in Electrical Engineering and the M.S. degree in Electrical and Computer Engineering from the Indian Institute of Technology- Bombay and University of Michigan - Ann Arbor in 1988 and 1992, respectively. He received the Ph.D. degree in Electrical and Computer Engineering from the University of Maryland - College Park in 1999. His research interests are in the areas of wireless and sensor network security, control and security of cyber-physical systems, adversarial modeling, smart connected communities, control-security, games-security, and information theoretic security in the context of wireless mobile networks. He is a Fellow of the IEEE for his contributions to security in cyber-physical systems. He is a recipient of the NSA LUCITE Rising Star Award (1999), National Science Foundation CAREER (2001), ARO YIP (2002), ONR YIP (2004), and PECASE (2005) for his research contributions to multi-user wireless security. He was co-author of award-winning papers including IEEE/IFIP William C. Carter Award Paper (2010) and WiOpt Best Paper Award (2012).
    \end{IEEEbiography}
    

\begin{thebibliography}{1}
\bibliographystyle{IEEEtran}

\bibitem{energy}
IRENA (2019), Global energy transformation: A roadmap to 2050 (2019 edition), International Renewable Energy Agency, Abu Dhabi.

\bibitem{battery}
“Future electric vehicles market analysis,” Accessed: Jan 27, 2021. [Online].
Available: https://www.alliedmarketresearch.com/electric-vehiclemarket

\bibitem{plugged-in}
G. Li and X. Zhang, "Modeling of Plug-in Hybrid Electric Vehicle Charging Demand in Probabilistic Power Flow Calculations," in IEEE Transactions on Smart Grid, vol. 3, no. 1, pp. 492-499, March 2012.

\bibitem{pareek}
S. Pareek, A. Sujil, S. Ratra and R. Kumar, "Electric Vehicle Charging Station Challenges and Opportunities: A Future Perspective," 2020 International Conference on Emerging Trends in Communication, Control and Computing (ICONC3), 2020, pp. 1-6, doi: 10.1109/ICONC345789.2020.9117473.

\bibitem{suh-kim}
I. Suh and J. Kim, "Electric vehicle on-road dynamic charging system with wireless power transfer technology," 2013 International Electric Machines \& Drives Conference, 2013, pp. 234-240, doi: 10.1109/IEMDC.2013.6556258.


\bibitem{olev}
J. Shin et al., "Contactless power transfer systems for On-Line Electric Vehicle (OLEV)," 2012 IEEE International Electric Vehicle Conference, 2012, pp. 1-4, doi: 10.1109/IEVC.2012.6183255.

\bibitem{machura-critical}
Philip Machura, Quan Li, A critical review on wireless charging for electric vehicles, Renewable and Sustainable Energy Reviews, Volume 104, 2019, Pages 209-234, ISSN 1364-0321, https://doi.org/10.1016/j.rser.2019.01.027.

\bibitem{hutchinson}
Hutchinson, Luke \& Waterson, Ben \& Anvari, Bani \& Naberezhnykh, Denis. (2019). Potential of Wireless Power Transfer for Dynamic Charging of Electric Vehicles. IET Intelligent Transport Systems. 13. 3-12. 10.1049/iet-its.2018.5221. 

\bibitem{road-example}
Laporte, S.; Coquery, G.; Deniau, V.; De Bernardinis, A.; Hautière, N. Dynamic Wireless Power Transfer Charging Infrastructure for Future EVs: From Experimental Track to Real Circulated Roads Demonstrations. World Electr. Veh. J. 2019, 10, 84. https://doi.org/10.3390/wevj10040084

\bibitem{boneh-franklin}
Dan Boneh and Matt Franklin, Identity-Based Encryption from the {Weil} Pairing, SIAM J. of Computing, Vol. 32, No. 3, pp. 586-615, 2003, Extended abstract in Crypto 2001, LNCS 2139, pp. 213-229, 2001.

\bibitem{hussain2015}
R. Hussain, D. Kim, M. Nogueira, J. Son, A. Tokuta and H. Oh, "A New Privacy-Aware Mutual Authentication Mechanism for Charging-on-the-Move in Online Electric Vehicles," 2015 11th International Conference on Mobile Ad-hoc and Sensor Networks (MSN), 2015, pp. 108-115, doi: 10.1109/MSN.2015.31.

\bibitem{li}
H. Li, G. Dán and K. Nahrstedt, "Portunes+: Privacy-Preserving Fast Authentication for Dynamic Electric Vehicle Charging," in IEEE Transactions on Smart Grid, vol. 8, no. 5, pp. 2305-2313, Sept. 2017, doi: 10.1109/TSG.2016.2522379.

\bibitem{hussain2017}
Rasheed Hussain, Junggab Son, Donghyun Kim, Michele Nogueira, Heekuck Oh, Alade O. Tokuta, Jungtaek Seo, "PBF: A New Privacy-Aware Billing Framework for Online Electric Vehicles with Bidirectional Auditability", Wireless Communications and Mobile Computing, vol. 2017, Article ID 5676030, 17 pages, 2017. https://doi.org/10.1155/2017/5676030

\bibitem{zhao}
Xingwen Zhao, Jiaping Lin, Hui Li, "Privacy-Preserving Billing Scheme against Free-Riders for Wireless Charging Electric Vehicles", Mobile Information Systems, vol. 2017, Article ID 1325698, 9 pages, 2017. https://doi.org/10.1155/2017/1325698

\bibitem{rabieh}
K. Rabieh and M. Wei, "Efficient and privacy-aware authentication scheme for EVs pre-paid wireless charging services," 2017 IEEE International Conference on Communications (ICC), 2017, pp. 1-6, doi: 10.1109/ICC.2017.7996868.

\bibitem{roman}
Roman, Luis \& Gondim, Paulo. (2019). Authentication Protocol in CTNs for a CWD-WPT Charging System in a Cloud Environment. Ad Hoc Networks. 97. 102004. 10.1016/j.adhoc.2019.102004. 

\bibitem{hamouid}
K. Hamouid and K. Adi, "Privacy-aware Authentication Scheme for Electric Vehicle In-motion Wireless Charging," 2020 International Symposium on Networks, Computers and Communications (ISNCC), 2020, pp. 1-6, doi: 10.1109/ISNCC49221.2020.9297199.

\bibitem{elghanam}
Elghanam, Eiman \& Ahmed, Ibtihal \& Hassan, Mohamed \& Osman, Ahmed. (2021). Authentication and Billing for Dynamic Wireless EV Charging in an Internet of Electric Vehicles. Future Internet. 13. 257. 10.3390/fi13100257. 

\bibitem{babu}
P. R. Babu, R. Amin, A. G. Reddy, A. K. Das, W. Susilo and Y. Park, "Robust Authentication Protocol for Dynamic Charging System of Electric Vehicles," in IEEE Transactions on Vehicular Technology, vol. 70, no. 11, pp. 11338-11351, Nov. 2021, doi: 10.1109/TVT.2021.3116279.

\bibitem{charm}
Charm-Crypto Library, jhuisi.github.io/charm/index.html

\bibitem{wan}
Wan, Zhiguo \& Ren, Kaili \& Preneel, Bart. (2008). A secure privacy-preserving roaming protocol based on hierarchical identity-based encryption for mobile networks. WiSec'08: Proceedings of the 1st ACM Conference on Wireless Network Security. 62-67. 10.1145/1352533.1352544. 

\bibitem{Bellare}
M. Bellare, A. Desai, E. Jokipii and P. Rogaway, "A concrete security treatment of symmetric encryption," Proceedings 38th Annual Symposium on Foundations of Computer Science, 1997, pp. 394-403, doi: 10.1109/SFCS.1997.646128.

\bibitem{Abdalla}
Michel Abdalla, Pierre-Alain Fouque, and David Pointcheval. 2005. Password-Based authenticated key exchange in the three-party setting. In Proceedings of the 8th international conference on Theory and Practice in Public Key Cryptography (PKC'05). Springer-Verlag, Berlin, Heidelberg, 65–84. DOI:https://doi.org/10.1007/978-3-540-30580-4\_6


\bibitem{ban}
Michael Burrows and Martín Abadi and Roger Needham. A logic of authentication. ACM Transaction on Computer Systems (1990). Volume 8, pp. 18-36.

\bibitem{stride1}
A. I. Swapna, M. R. Huda and M. K. Aion, "Comparative security analysis of software defined wireless networking (SDWN)-BGP and NETCONF protocols," 2016 19th International Conference on Computer and Information Technology (ICCIT), 2016, pp. 282-287.

\bibitem{stride2}
F. Ruffy, W. Hommel and F. Von Eye, "A STRIDE-based Security Architecture for Software- Defined Networking", The Fifteenth International Conference on Networks (ICN), 2016.

\bibitem{stride}
M. Howard and S. Hernan. Uncover security design flaws using the stride approach. (Accessed on 09/05/2016). [Online]. Available: https://msdn.microsoft.com/magazine/msdn-magazine-issues

\bibitem{dolev-yao}
D. Dolev and A. Yao, "On the security of public key protocols," in IEEE Transactions on Information Theory, vol. 29, no. 2, pp. 198-208, March 1983.

\end{thebibliography}
\end{document}